\documentclass[prl,aps,reprint,twocolumn,showpacs,superscriptaddress]{revtex4-1}
\usepackage{graphicx,color,bm}
\usepackage[english]{babel}
\usepackage{amsmath,amssymb}
\usepackage{dsfont}
\usepackage{subfigure}
\newcommand{\magent}[1]{{\textcolor{black}{#1}}}
\newcommand{\green}[1]{{\textcolor{black}{#1}}}
\newcommand{\red}[1]{{\textcolor{black}{#1}}}
\newcommand{\cyan}[1]{{\textcolor{black}{#1}}}
\newcommand{\blue}[1]{{\textcolor{black}{#1}}}

\begin{document}

\title{Persistent Skyrmion Lattice of \red{Noninteracting} Electrons with Spin-Orbit Coupling}
\author{Jiyong Fu}
\thanks{Permanent address: Department of Physics, Qufu Normal University, Qufu, Shandong, 273165, China.}
\affiliation{Instituto de F\'{\i}sica de S\~ao Carlos, Universidade de S\~ao Paulo, 13560-970, S\~ao Carlos, S\~ao Paulo, Brazil}
\author{Poliana H. Penteado}
\affiliation{Instituto de F\'{\i}sica de S\~ao Carlos, Universidade de S\~ao Paulo, 13560-970, S\~ao Carlos, S\~ao Paulo, Brazil}
\affiliation{Department of Physics and Astronomy, University of California, Los Angeles, California 90095, USA}
\author{Marco O. Hachiya}
\affiliation{Instituto de F\'{\i}sica de S\~ao Carlos, Universidade de S\~ao Paulo, 13560-970, S\~ao Carlos, S\~ao Paulo, Brazil}
\author{Daniel Loss}
\affiliation{Department of Physics, University of Basel, CH-4056 Basel, Switzerland}
\author{J. Carlos Egues}
\affiliation{Instituto de F\'{\i}sica de S\~ao Carlos, Universidade de S\~ao Paulo, 13560-970, S\~ao Carlos, S\~ao Paulo, Brazil}

\date{\today}

\begin{abstract}
A persistent spin helix (PSH) is a robust helical spin-density pattern arising in disordered 2D electron gases with Rashba $\alpha$ and Dresselhaus $\beta$ spin-orbit (SO) tuned couplings, i.e., $\alpha=\pm\beta$. Here we investigate the emergence of a Persistent Skyrmion Lattice (PSL) resulting from the coherent superposition of PSHs along orthogonal directions -- crossed PSHs -- in wells with two occupied subbands $\nu=1,2$.
For realistic GaAs wells we show that the Rashba $\alpha_\nu$ and Dresselhaus $\beta_\nu$ couplings can be simultaneously tuned to equal strengths but opposite signs, e.g., $\alpha_1= \beta_1$ and $\alpha_2=-\beta_2$. In this regime and away from band anticrossings, our {\it non-interacting} electron gas sustains a topologically non-trivial skyrmion-lattice spin-density excitation, which inherits the robustness against spin-independent disorder and interactions from its underlying crossed PSHs. We find that the spin relaxation rate due to the interband SO coupling is comparable to that of the cubic Dresselhaus term as a mechanism of the PSL decay. Near  anticrossings, the interband-induced spin mixing leads to unusual spin textures along the energy contours beyond those of the Rahsba-Dresselhaus bands. Our PSL opens up the unique possibility of observing topological phenomena, e.g., topological and skyrmion Hall effects, in ordinary GaAs wells with non-interacting electrons. 
\end{abstract}

\maketitle

Topological spin textures in crystals arise in connection with the electron-electron interaction. Skyrmions in the fractional quantum Hall regime~\cite{Sondhi,Brey}, magnetic and multiferroic systems \cite{skyrmions} exemplify spin patterns characterized by topological invariants associated with the nontrivial winding of the spins. Non-topological helical spin patterns, e.g., spin-density waves in metals~\cite{Overhauser} can also occur. When coupled to conduction electrons, the emergent electrodynamics of the non-trivial spin textures gives rise to fundamental phenomena, e.g., the topological {\it and} skyrmion Hall effects in chiral magnets~\cite{naga-toku}.

\begin{figure}[t!]
\centerline{\resizebox{2.8992in}{!}{\includegraphics{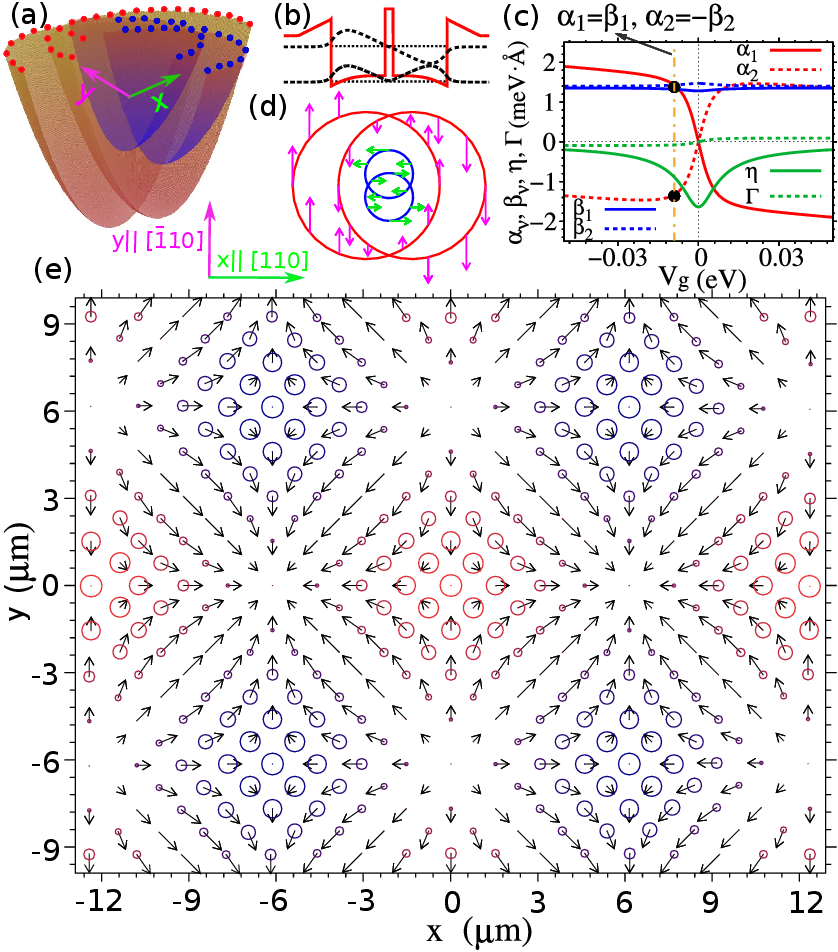}}}
\caption{\label{fig1} (Color online)
(a) Energy dispersion for a GaAs double well with two subbands (no disorder) and (b) its potential profile and wave functions. (c) Calculated SO couplings vs. $V_g$: intraband (interband) Rashba $\alpha_\nu$ ($\eta$) and Dresselhaus $\beta_\nu$ ($\Gamma$).
The dot-dashed vertical line (orange) indicates the crossed PSH symmetry point $\alpha_1=\beta_1$ and $\alpha_2=-\beta_2$.
(d) Energy contours; the arrows pointing along the orthogonal axes $x$ (pink) and $y$ (green) define the subband SO fields within subband 1 and 2, respectively. (e) Persistent Skyrmion Lattice (PSL) pattern  in the 2DEG. The size of the circles and arrows denote $\langle s_z\rangle$ and $\langle s_{x,y}\rangle$, respectively. Blue (dark gray) circles stand for spins up and red (light gray) for spins down.}
\end{figure}

Here we show that non-interacting 2D electrons in two-subband quantum wells \cite{Bernardes,souma:2013} with {\it matched} SO couplings of opposite signs $\alpha_1=\beta_1>0$, $\alpha_2=-\beta_2<0$, can sustain a {\it Persistent Skyrmion Lattice} (PSL), Fig.~\ref{fig1}. This should allow the observation of fundamental topological phenomena in ordinary (non-magnetic) GaAs  wells.

The formation of this skyrmion lattice can be easily understood for ballistic electrons (later on we include disorder). For a single-subband well with $\alpha_1=\beta_1$, the Rashba-Dresselhaus Hamiltoninan is effectively 1D: $H_{RD1}= 2\alpha_1\sigma_y p_x/\hbar = g\mu_B \sigma_y B_y/2$, i.e., an electron-momentum ($p_x$)-dependent Zeeman interaction with a unidirectional effective magnetic field $B_y$ ($y \parallel [\bar 1 1 0]$, $x \parallel [110]$). Here $\sigma_y$ is the Pauli matrix and $\mu_B$ the Bohr magneton. The corresponding quantum evolution operator is: $U_{RD1}(t)=e^{-i  g\mu_B \sigma_y B_yt/2\hbar}=e^{-i\sigma_y Q_1 x/2}$, where $x=p_xt/m^*$, $Q_1=4m^*\alpha_1/\hbar^2$, and $m^*$ the electron mass. Hence  a spin-up electron injected at $x=y=0$ precesses around this $B_y$ field such that 
\begin{equation}
U_{RD1}\left(
\begin{array}{c}
1 \\
0
\end{array}
\right) = \left(
\begin{array}{c}
\cos (Q_1x/2) \\
\sin(Q_1x/2)
\end{array}
\right).
\label{spinor2x1}
\end{equation}
This spinor leads to  a spin-density wave in the first subband $\mathbf{s}^{1}(\mathbf{r})$, with $s_x^{1}\propto \sin (Q_1 x)$, $s_y^{1} = 0 $, $s_z^{1} \propto \cos(Q_1 x)$, and pitch $Q_1$. This helical pattern also arises in the presence of spin-independent disorder and (time-reversal conserving) interactions, and is  known as Persistent Spin Helix (PSH)~\cite{Schliemann, Bernevig-PSH} (see Ref.~\cite{ballistic-PSH} for `ballistic PSHs'). Koralek {\it et al.} first observed a PSH via transient spin grating spectroscopy~\cite{Koralek}; Walser {\it et al.} imaged PSHs using time-resolved Kerr rotation microscopy~\cite{Walser}. 
A single PSH is, however, non-topological. 

By considering a second subband with $\alpha_2=-\beta_2<0$ (Fig.~1), we can generalize Eq.~(\ref{spinor2x1}) so that a spin up electron injected into  both subbands evolves to
\begin{equation}
U_{RD1}\otimes U_{RD2}\left(
\begin{array}{c}
1 \\
0 \\
1 \\
0
\end{array}
\right) \rightarrow \left(
\begin{array}{c}
\cos (Q_1x/2) \\
\sin(Q_1x/2) \\
\cos(Q_2y/2)\\
i\sin(Q_2y/2)
\end{array}
\right), 
\label{spinor4x1}
\end{equation}
where $U_{RD2}=e^{i\sigma_xQ_{2} y/2}$. Here the second subband gives rise to a PSH with pitch $Q_2=4m^*\alpha_2/\hbar^2$ and spin density $s_x^{2} \propto 0$, $s_y^{2}\propto \sin(Q_{2}y)$, and $s_z^{2} \propto \cos(Q_{2}y)$, {\it orthogonal} to that of the first subband. \blue{These  crossed PSHs form the unconventional pattern $\propto \mathbf{s}^{1}(\mathbf{r}) + \mathbf{s}^{2}(\mathbf{r})$ in Fig.~1(e): a Persistent Skyrmion Lattice (PSL), \red{that} shows regions of zero and max/min spin densities characterized by a
topological invariant (skyrmion number). } 

The PSL texture inherits the robustness of the crossed spin helices, which are protected by the underlying $SU(2)$ symmetry (within each subband) in lowest order of the cubic and interband SO interactions. More physically, this robustness follows from the partial cancellation of the linear-in-momentum Rashba and Dresselhaus SO terms for $\alpha_1=\beta_1$ and $\alpha_2=-\beta_2$, which renders unidirectional SO fields within each subband [Fig.~1(d)], and underlies the emergence of spin-conserved quantities in the system~\cite{Schliemann,Bernevig-PSH}. As the electrons move, they undergo spin rotations about orthogonal effective magnetic fields thus forming the skyrmion pattern in Fig.~1(e). Note that our PSL is identical to the `spin crystal' of Ref.~\cite{binz}. 

\begin{figure}[t]
\centerline{\resizebox{3.3in}{!}{
\includegraphics{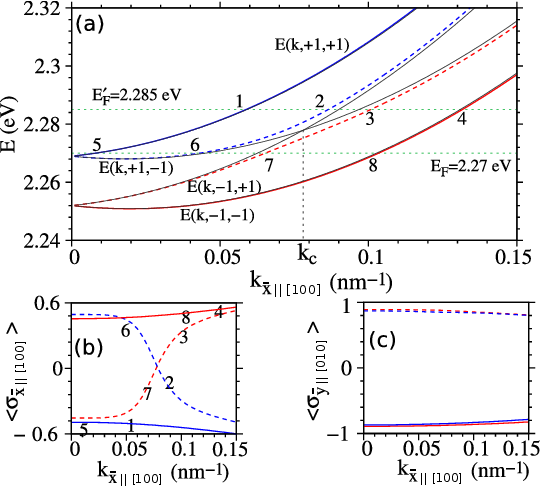}}}
\caption{(Color online) (a) Energy dispersions $E_{\mathbf{k},\lambda_1,\lambda_2}$ (scaled by a factor of 10 for visibility) along $k_{\bar x} \parallel$ [100] (or  $k_{x}=k_{y}$) for an InSb double well.
The black solid lines correspond to the uncoupled ($\eta=\Gamma=0$) bands $E_{\nu,\mathbf{k}}^{\pm}$ and cross at $k_c$. For $\eta$, $\Gamma\neq0$ these bands anticross (dashed lines). Away from $k_c$, the coupled and uncoupled cases coincide. The label sets (1, 2, 3, 4)  and (5, 6, 7, 8)  denote Fermi points along $k_{\bar x}$ at $E_{F}^\prime$ and $E_F$, respectively. Panels (b) and (c) show $\langle \sigma_{\bar x} \rangle$ and $\langle \sigma_{\bar y} \rangle$, respectively, along $k_{\bar x}$.
The solid and dashed lines correspond to the respective energy branches in~(a).} 
\label{fig2}
\end{figure}

We have also derived analytical expressions for the PSL spin density [Fig.~\ref{fig1}(e)] in the presence of disorder both (i) quantum mechanically~\cite{Schliemann} and (ii) via diffusive  equations~\blue{\cite{Bernevig-PSH,Mishchenko,Raimondi2006,Duckheim,Wang,Sinova,Shen2014}}. We show that intersubband-induced spin relaxation limits the PSL lifetime similarly (in magnitude) to the cubic Dresselhaus term; PSLs are then  feasible. With no disorder, our energy dispersions feature two Dirac cones at $\mathbf{k}=0$ Fig.~\ref{fig1}(a), an  anticrossing with spin mixing [significant in InSb, Fig.~\ref{fig2}(a)], and highly anisotropic four-branch Fermi contours Fig.~\ref{fig1}(d).

\paragraph*{Model Hamiltonian.---} We consider a quantum well with two subbands. The two lowest spin-degenerate eigensolutions are $\langle \mathbf{r}| \mathbf{k},\nu,\sigma\rangle=e^{i\mathbf{k}\cdot\mathbf{r}}\varphi_\nu(z)|\sigma_z\rangle$, $\nu=1,2$ and $\sigma_z = \uparrow, \downarrow$, with energies $\varepsilon_{\nu,k} = \varepsilon_{\nu} + \hbar^2k^2/2m^*$, where $\mathbf{k}$ is the in-plane electron wave vector and $\varepsilon_\nu$ is the $\nu$th confined well level. Here we  generalize the usual single-subband Rashba-Dresselhaus Hamiltonian for this two-subband case,  which reads to linear order in $k$ \magent{[see Supplemental Material (SM), Sec.~(I.A), for details \red{\cite{suppl-mat-refs}}]}
\begin{widetext}
\begin{eqnarray}
\tilde{\mathcal{H}} = \left(
       \begin{array}{cc}
	 \varepsilon_{1,k} \mathds{1}+\alpha_1(\sigma_{\bar{y}} k_{\bar{x}}-\sigma_{\bar{x}}k_{\bar{y}}) + \beta_1(\sigma_{\bar{y}} k_{\bar{y}}-\sigma_{\bar{x}} k_{\bar{x}})     & \eta(\sigma_{\bar{y}} k_{\bar{x}}-\sigma_{\bar{x}}k_{\bar{y}}) + \Gamma(\sigma_{\bar{y}} k_{\bar{y}}-\sigma_{\bar{x}} k_{\bar{x}})   \\
	  \eta(\sigma_{\bar{y}} k_{\bar{x}}-\sigma_{\bar{x}}k_{\bar{y}}) + \Gamma(\sigma_{\bar{y}} k_{\bar{y}}-\sigma_{\bar{x}} k_{\bar{x}})  & \varepsilon_{2,k} \mathds{1} +\alpha_2(\sigma_{\bar{y}} k_{\bar{x}}-\sigma_{\bar{x}}k_{\bar{y}}) + \beta_2(\sigma_{\bar{y}} k_{\bar{y}}-\sigma_{\bar{x}} k_{\bar{x}}) 
       \end{array}
       \right),    \label{hamil-subbands1}
\end{eqnarray}
\end{widetext}
where $\sigma_{\bar{x},\bar{y}}$ are the spin Pauli matrices, $k_{\bar{x},\bar{y}}$ the wave vector components along the $\bar{x} \parallel[100]$ and $\bar{y} \parallel[010]$ directions, and $\alpha_\nu$, $\beta_\nu$, the Rashba and Dresselhaus {\it intrasubband} couplings, respectively, for subbands $\nu=1,2$. Note that Eq.~(\ref{hamil-subbands1}) accounts for SO-induced {\it intersubband} couplings \cite{Calsaverini,Bernardes} via the parameters $\eta$ (Rashba) and $\Gamma$ (Dresselhaus)~\cite{real-param}. Note that $\tilde{\mathcal{H}}$ describes two usual  Rashba-Dresselhaus systems [the $2\times2$ upper left ($\alpha_1$, $\beta_1$) and lower right ($\alpha_2$, $ \beta_2$) blocks] coupled via the intersubband ``off-diagonal blocks" ($\eta$, $\Gamma$). The energy dispersions of $\tilde{\mathcal{H}}$ display anticrossing near $k_c$, e.g., Fig.~2(a) for InSb wells. Similar dispersions (not shown) hold for a GaAs. As we show in the SM [Sec. (II.C)], for typical electron densities the Fermi wave vectors are such that $k_F << k_c$ for GaAs and $k_F \sim k_c$ for InSb wells. Next we use L\"owdin perturbation theory to decouple (in orders of $k$ or the subband energy separation $\Delta\varepsilon$) the two Rashba-Dresselhaus blocks in Eq.~(\ref{hamil-subbands1}); this procedure is valid for $k= k_F<<k_c$ as we discuss in the SM [Sec~(I)].

For convenience, let us first rotate the axes around $z$ such that $\bar{x} \rightarrow x \parallel[110]$, $\bar{y}\rightarrow y \parallel[\bar{1}10]$ ($z\parallel [001]$) and then perform a spin rotation $R=e^{-i\sigma_z \theta_z/2}$, with $\theta_z=\pi/4$. 
\magent{To lowest order, we find the uncoupled blocks}  
\begin{equation}
 \mathcal{H}_\nu=\varepsilon_{\nu,k} \mathds{1}+(-\alpha_\nu+\tilde{\beta}^{1 y}_\nu) \sigma_{x}k_{y} +(\alpha_\nu+\tilde{\beta}^{1x}_\nu) \sigma_{y} k_{x}, 
 \label{subHs}
 \end{equation}
\green{in which ${m^*}$ is the effective mass, $\tilde{\beta}^{1i}_\nu = \beta^1_{\nu} - \beta^3_{\nu} - \beta^{1i}_\nu$, with $i=x,y$; $ \beta^1_{\nu} = \gamma \langle \nu | k_z^2 | \nu \rangle$ (``bare'' linear Dresselhaus), $\beta^3_{\nu}\simeq \gamma \pi n_\nu /2$,  $n_\nu$ the subband areal density and $\gamma=11.0$ $\text{eV\AA}^3$ the bulk Dresselhaus constant \cite{Dominik} and $\beta^{1i}_\nu$ is a function of $\Delta \varepsilon = \varepsilon_2-\varepsilon_1$, \magent{$n_\nu$}, $\eta$ and $\Gamma$ [Eqs.~(S12)-(S13), SM].} \green{As we show in the SM [see discussion after Eq.~(S26)], $\beta^{1i}_\nu<<\beta^3_{\nu}$; hence in what follows we take the couplings  $\tilde{\beta}^{1i}_\nu$ in Eq.~(\ref{subHs}) to be  $\beta_\nu=\beta^1_{\nu} - \beta^3_{\nu}$. This shows that the intersubband couplings ($\eta$, $\Gamma$) essentially do not alter the PSH condition within each subband,  $\alpha_\nu=\pm \beta_\nu$.} 


Equation (\ref{subHs}) shows that our two-subband well can be described (to linear order in k) as two uncoupled ``copies'' of the usual single-subband Rashba-Dresselhaus model $ \mathcal{H}_\nu$, with renormalized parameters. Each copy has SU(2) symmetry at $\alpha_\nu=\pm \beta_\nu$~\cite{Bernevig-PSH}. Next we show that the unique matching $\alpha_1=\beta_1>0$ and $\alpha_2=-\beta_2<0$ occurs in realistic GaAs wells \cyan{[Fig.~1(a) shows the energy dispersions $E_{\nu,\mathbf{k}}^{\pm}$ in this case, see Eq.~(S27) in the SM]}.

{\it SO couplings for GaAs wells.---} We self-consistently solve both Schr\"odinger and Poisson's equations to obtain the eigenfunctions $\varphi_\nu(z)$ of a GaAs double well (similar results hold for a single wide well), Fig.~\ref{fig1}(b)\magent{\red{;} see SM, Sec.~(II)}. From $\varphi_\nu(z)$ we calculate the SO couplings, \cyan{Fig.~\ref{fig1}(c). 
While $\alpha_1$ and $\alpha_2$} have opposite signs and are very sensitive to the gate voltage $V_g$ across the well, $\beta_1$ and $\beta_2$ ($\beta_1 \approx \beta_2$) are practically constant~\cite{Calsaverini}.  At $V_g=-10$~meV (dot-dashed vertical line), we find $\alpha_1=\beta_1=-\alpha_2=\beta_2=1.45$ $\text{meV \AA}$ (black dots), which enables crossed PSHs as we discuss below. 

\paragraph*{Robust eigenspinors even with disorder. --- }  More realistically, 
we \magent{now} consider an arbitrary time-reversal symmetric potential $V(\mathbf{r})$ \magent{(spin independent)}, e.g., due to non-magnetic disorder~\cite{Schliemann}, in Eq.~(\ref{subHs}): $\mathcal{H}^\nu_{dis}=\mathcal{H}_\nu+V(\mathbf{r})$.  For \magent{$\alpha_1 = \beta_1>0$ and $\alpha_2=-\beta_2<0$} [Fig.~1(c)], $\mathcal{H}^\nu_{dis}$ admits eigenstates of the form
$\psi_1^{\uparrow_{y}}(\mathbf{r}) = \varphi(\mathbf{r})e^{-iQ_1 x/2}|\uparrow_{y}\rangle$ and 
$\psi_1^{\downarrow_{y}}(\mathbf{r}) = \varphi(\mathbf{r})e^{iQ_1 x/2}|\downarrow_{y}\rangle$ for subband 1, and
$\psi_2^{\uparrow_{x}}(\mathbf{r}) = \varphi(\mathbf{r})e^{iQ_2 y/2}|\uparrow_{x}\rangle$ and 
$\psi_2^{\downarrow_{x}}(\mathbf{r}) = \varphi(\mathbf{r})e^{-i Q_2 y/2}|\downarrow_{x}\rangle$ for subband 2. Here $Q_\nu=4m^*\alpha_\nu/\hbar^2$ ($\nu=1,2$, \magent{$Q_1>0$, $Q_2<0$)} and $|\uparrow_{x}\rangle$, $|\downarrow_{x}\rangle$ ($|\uparrow_{y}\rangle$, $|\downarrow_{y}\rangle$) are the eigenvectors of $\sigma_{x}$ ($\sigma_y$). The ``envelope function'' $\varphi(\mathbf{r})$ satisfies 
$\left(-\hbar^2\nabla^2/2m^* + V(\mathbf{r})\right)\varphi(\mathbf{r})=\left(\varepsilon - \varepsilon_\nu + 2\alpha_\nu^2m^*/\hbar^2\right)\varphi(\mathbf{r})$ \cite{note-pot}. Because $[\mathcal{H}^1_{dis}, \sigma_{y}]=0$ and $[\mathcal{H}^2_{dis}, \sigma_{x}]=0$, the eigensolutions of $\mathcal{H}^\nu_{dis}$ possess (i) robust spin states against \magent{non-magnetic scattering} (the spin and orbital variables factorize in $\mathcal{H}^\nu_{dis}$), and (ii) definite SO-induced spin-rotation phases dependent only on the distance `traveled' along \magent{$x$ and $y$}, respectively. From (i) and (ii) we construct next ``skyrmion states".

\paragraph*{\hbox{Persistent Skyrmion Lattice: Quantum approach. ---}} Let $\psi_1(\mathbf{r})=\varphi(\mathbf{r})\left(e^{-i Q_1 x/2}|\uparrow_{y}\rangle + e^{i Q_1 x/2}|\downarrow_{y}\rangle\right)/\sqrt{2}$ and $\psi_2(\mathbf{r})=\varphi(\mathbf{r})\left(e^{iQ_2 y/2}|\uparrow_{x}\rangle + e^{-iQ_2 y/2}|\downarrow_{x}\rangle\right)/\sqrt{2}$, i.e.,  stationary spin up states at $\mathbf{r}=0$ for each subband $\nu=1,2$ at the Fermi energy $E_F$. The corresponding spin densities $\mathbf{s}^{\nu}(\mathbf{r})$ within each subband are: $\mathbf{s}^{1}(\mathbf{r}) = \frac{1}{2}\psi^\dagger_1(\mathbf{r}) \boldsymbol{\sigma} \psi_1(\mathbf{r}) \propto \sin (Q_1 x)\hat {x} + \cos (Q_1 x)\hat {z}$ and $\mathbf{s}^{2}(\mathbf{r}) = \frac{1}{2}\psi^\dagger_2(\mathbf{r}) \boldsymbol{\sigma} \psi_2(\mathbf{r}) \propto \sin (Q_2 y)\hat {y} + \cos (Q_2 y)\hat {z}$, with $\boldsymbol{\sigma}=\sigma_{x} \hat {x} + \sigma_{y} \hat {y} + \sigma_z \hat z$. These are orthogonal PSHs. Considering now the stationary superposition $\psi(\mathbf{r})=[\psi_1(\mathbf{r}) \oplus \psi_2(\mathbf{r})]/\sqrt{2}$  at $E_F$ (this is feasible \magent{\cite{Koralek,Walser}} as we discuss later on), we can calculate 
the spin density $\mathbf{s}(\mathbf{r})=
\psi^\dagger (\mathbf{r})( \mathds{1}\otimes \frac{1}{2} \boldsymbol{\sigma}) \psi (\mathbf{r})$ (in units of $\hbar$) 
\begin{eqnarray}
\mathbf{s}(x,y) &=&\dfrac{1}{4}|\varphi(\mathbf{r})|^2\Big\{\sin (Q_1 x)\hat {x} +  \sin (Q_2 y) \hat {y} +  \nonumber \\  && \big[\cos (Q_1 x) + \cos (Q_2 y) \big]\hat {z}\Big\}.
\label{expectation}
\end{eqnarray}
Interestingly, $\mathbf{s}(\mathbf{r})=(\mathbf{s}^{1}(\mathbf{r}) +\mathbf{s}^{2}(\mathbf{r}))/2 $ forms a `persistent skyrmion lattice'  \cite{binz}, Fig.~\ref{fig1}(e), arising  from  two orthogonal PSHs, along $\hat x$ (1st subband) and $\hat y$ (2nd subband). We assume $\varphi(\mathbf{r}) \simeq e^{i\mathbf{k}\cdot\mathbf{r}}$  (``weak disorder''), i.e., $|\varphi(\mathbf{r})|^2 = 1$ in Eq.~(\ref{expectation}), to obtain Fig.~\ref{fig1}(e). Our  PSL inherits the robustness from its constituent persistent spin helices. In analogy to Ref.~\cite{binz}, we can define $\hat{\mathbf{n}}= \mathbf{s}/|\mathbf{s}|$ and show that the PSL is characterized by a skyrmion number over its unit cell area $S$: $\frac{1}{4\pi}\int_S \hat{\mathbf{n}}\cdot(\partial_x \hat{\mathbf{n}}\times \partial_y \hat{\mathbf{n}}) dxdy$. Next we corroborate the quantum results presented here via diffusive equations.

\paragraph*{Semiclassical approach. ---} Following Refs.~\cite{Sinova,Shen2014}, we solve a set of diffusive transport equations for the coupled dynamics of charge $n^\nu(x,y,t)$ and spin 
$s^\nu_{x,y,z}(x,y,t)$ densities in subbands $\nu=1,2$, valid in the weak SO interaction limit \cyan{$(\alpha_\nu) \beta_\nu k_F \tau/\hbar<<1$}, $\tau$ being the momentum scattering time [see SM, Sec.~(III), for details]. 

At $\alpha_1=\beta_1$ and $\alpha_2=-\beta_2$ (symmetry point), the Fourier components of the spin density $s^{\nu}_{j}(q_x,q_y,t)$ obey
\begin{eqnarray}
s_{j}^{\nu}(\mathbf{q},t) = A^\nu_{j,+}(\mathbf{q})e^{\omega^{\nu}_{j,+}(\mathbf{q})t} + A^{\nu}_{j,-}(\mathbf{q})e^{\omega^{\nu}_{j,-}(\mathbf{q})t}, 
\label{szq}
\end{eqnarray}
where $A^{\nu}_{j,{\pm}}(\mathbf{q})$, $j=x, y, z$, are amplitudes set by the initial conditions, $\omega^1_{z,\pm}=\omega^1_{x,\pm}=-Dq^2-T^{1}_x \pm C^1_x q_x\equiv \omega^1_{\pm}$, 
$\omega^2_{z,\pm}=\omega^2_{y,\pm}=-Dq^2-T^{2}_y \pm C^2_y q_y\equiv \omega^2_{\pm}$,  and $\omega^1_{y,\pm}=\omega^2_{x,\pm}=-Dq^2$, with $D=v_F^2\tau/2$ the diffusion constant, $v_F=\hbar k_F/m^*$ the Fermi velocity, $q^2=q_x^2+q_y^2$, $C^1_x=4\alpha_1 k_F^2\tau/m^*$, $C^2_y=-4\alpha_2 k_F^2\tau/m^*$, $T^1_x=8\alpha_1^2 k_F^2 \tau/\hbar^2$, and $T^2_y=8\alpha_2^2 k_F^2 \tau/\hbar^2$. 
Hence given an arbitrary initial spin density $\mathbf{s}(\mathbf{q},t=0)$ [or equivalently $\mathbf{s}(\mathbf{r},t=0)$], we can determine $s_j(\mathbf{q},t) = s_j^1(\mathbf{q},t) + s^2_j(\mathbf{q},t)$ and the spin-density profile $s_j(\mathbf{r},t) =\int  s_j(\mathbf{q},t) e^{-i \mathbf{q}.\mathbf{r}}d\mathbf{q}$ at any time $t$. Next we discuss how PSLs can be realized.

\paragraph*{Exciting a PSL via transient spin grating ---}   The setup in \cite{Koralek} can be implemented with crossed lasers of wave vectors $\mathbf{q}^a=q^a_{x}\hat x$ and $\mathbf{q}^b=q^b_{y}\hat y$. This creates orthogonal spin gratings with an initial spin density $s_z(\mathbf{r},0) \propto \cos{(q^a_{x} x)} + \cos{(q^b_{y} y)}$ and $s_{x}(\mathbf{r},0)=s_{y}(\mathbf{r},0)=0$. In real space, the resulting $z$-component $s_z^1 + s_z^2$ reads
\begin{eqnarray}
 s_{z}\left(\mathbf{r},t\right) &=& \dfrac{1}{4}\left(e^{\omega^1_{+}(\mathbf{q}^a)t} + e^{\omega^1_{-}(\mathbf{q}^a)t}+2e^{\omega^2_{+}(\mathbf{q}^a)t}  \right)\cos{(q^a_{x} x)}  \nonumber \\   
                                            &+& \dfrac{1}{4}\left(e^{\omega^2_{+}(\mathbf{q}^b)t} +  e^{\omega^2_{-}(\mathbf{q}^b)t}+  2e^{\omega^1_{+}(\mathbf{q}^b)t}\right) \cos{(q^b_{y} y)}. \nonumber \\
 \label{szr} 
\end{eqnarray}
Similarly we find for $x$ and $y$, respectively,
\begin{eqnarray}
&s_{x}\left(\mathbf{r},t\right)&= \dfrac{1}{4}\left(e^{\omega^1_+(\mathbf{q}^a)t} - e^{\omega^1_-(\mathbf{q}^a)t} \right)\sin{(q^a_{x} x)}, \label{sxp}  \\
&s_{y}\left(\mathbf{r},t\right)&=-\dfrac{1}{4}\left(e^{\omega^2_+(\mathbf{q}^b)t} - e^{\omega^2_-(\mathbf{q}^b)t} \right)\sin{(q^b_{y} y)}.
\label{sxm}
\end{eqnarray}
The spin grating experiment of \cite{Koralek} uses one laser  and finds two decay constants for $s_z$ in a one-subband well [Eq.~(\ref{szq})]. Here we have two subbands and two lasers; hence we find four time constants in each subband,  \cyan{two of which are equal [Eq.~(\ref{szr})]; see also SM, Secs.~(III.B), (III.C)}. Equations~(\ref{szr})--(\ref{sxm}) show that the spin density excitation decays to zero as $t\rightarrow \infty$ for arbitrary $q_x^a$ and $q_y^b$. However, when the laser  \magent{wave vectors are tuned to match} the  pitches of the crossed PSHs, i.e., $\mathbf{q}^a=Q_1\hat x$,  ($\alpha_1=\beta_1$), and $\mathbf{q}^b=Q_2\hat y$, ($\alpha_2=-\beta_2$), we have $\omega^1_+(Q_1,0) =\omega^2_-(0,Q_2)=0$ and hence
$s_{x}\left(\mathbf{r},t\rightarrow \infty \right) = \sin{(Q_1x)/4}$, 
$s_{y}\left(\mathbf{r},t\rightarrow \infty \right) =\sin{(Q_2y)/4}$, and 
$s_z\left(\mathbf{r},t\rightarrow \infty \right) =\left[\cos{(Q_1 x)} + \cos{(Q_2 y)}\right]/4$.
This is the PSL within the diffusive approach (cf. the quantum result in Eq.~(\ref{expectation})).

\paragraph*{Self-forming PSL upon photo-excitation. ---} A single PSH evolves from a uniform photo-excited spin-polarized density (e.g., $s_z(\mathbf{r},0)=1$) in one-subband wells as demonstrated in \cite{Walser}. By the same token, a PSL (crossed PSHs) will also emerge in this setting, provided that $\alpha_1=\beta_1$ and $\alpha_2=-\beta_2$. Essentially, all Fourier components of  $s_z(\mathbf{r},0)=1$ decay to zero as $t\rightarrow \infty$, except those  with the two ``magic" $\mathbf{q}$'s: $\mathbf{q_1}=Q_1\hat x$ and $\mathbf{q_2}=Q_2\hat y$, thus leading to crossed helices or a PSL [see SM, Sec.~(III.D)].


{\em Detrimental effects to the PSL}--- Our PSL so far has an infinite lifetime. It is known that in single-subband wells the cubic Dresselhaus \magent{(neglected so far)} limits the lifetime of persistent spin helices. In addition, our PSL arises in two-subband wells and inter-subband spin decay may be an issue. However, we show that the spin relaxation rate due to the interband SO coupling (Elliott-Yafet type)~\cite{Oestreich} is comparable to that of the cubic Dresselhaus (D'yakonov-Perel type)~\cite{Bernevig-PSH,Koralek,luffe:2011} in limiting the PSL lifetime \magent{[see SM, Sec. (IV)]}. Furthermore, deviations from the PSL condition $\alpha_1=\beta_1$ (and/or $\alpha_2=-\beta_2$) such that $\alpha_\nu = \alpha_\nu + \delta_\nu$, with $|\delta_\nu/\alpha_\nu|\ll 1$ ($\nu=1,2$), induce spin scattering with (golden-rule) rates  $\sim\delta_\nu^2$, i.e., spin dephasing vanishes in linear order in $\delta_\nu$ \cite{Schliemann,kiselev}. Hence PSLs should be feasible with the setups in~\cite{Koralek,Walser}.

To mitigate the stringency of the ``$\alpha=\beta$'' condition at a unique value, we note that both $\alpha$ and $\beta$ can be varied {\it simultaneously for a single sample} -- while still keeping $\alpha=\beta$ -- over a wide range of electron densities in single-subband GaAs wells as shown in Ref.~\cite{Dominik}. This allows for helices with gate-tunable pitches and ultimately to skyrmion lattices with controllable lattice constants (provided the findings in \cite{Dominik} hold for two-subband wells).

\paragraph*{Band anticrossing $\&$ spin texture in k space.---} We now turn to the effects of the interband couplings $\eta$ and $\Gamma$ on the energy spectrum of $\tilde{\mathcal{H}}$ [Eq.~(\ref{hamil-subbands1}), no disorder]. The solid lines in Fig.~\ref{fig2}(a) show the  bands $E_{\nu,\mathbf{k}}^{\pm}$ for the uncoupled case $\Gamma=\eta=0$ [see Eq.~(S27) in the SM~\cite{Schliemann}]. Both $\eta$ and $\Gamma$ couple these bands with distinct `spin' and orbital quantum numbers. 
Here we focus on InSb wells for which the SO coupling is stronger as compared with GaAs. The new bands $E_{\mathbf{k},\lambda_1,\lambda_2}$ ($\lambda_1, \lambda_2=\pm 1$) for non-zero $\Gamma$ and $\eta$ display anticrossings around $k_c$~\cite{kc-note}, Fig.~\ref{fig2}(a) (dashed lines), and 
a strong spin mixing in $\langle \sigma_{\bar x} \rangle$, Figs.~\ref{fig2}(b) and \ref{fig2}(c)~\cite{Bentmann}, near the anticrossing. 
This follows from an interplay of intersubband couplings: when either one of them is null, no spin mixing occurs as only same-spin branches couple in this case \cyan{[see Eq.~(S56) and discussion following it in the SM]. For completeness, in the SM we present the spin textures along the constant-energy contours $E_{\mathbf{k},\lambda_1,\lambda_2}=E_F$, $E_F^\prime$ [Fig.~\ref{fig2}(a)].} 

\paragraph*{Novel topological phenomena in 2DEGs?---} Similarly to chiral magnets, we conjecture that a PSL formed on top of an electrically drifting Fermi sea \cite{yang} can possibly lead to  topological~\cite{binz} and skyrmion~\cite{SKHE} Hall effects in ordinary GaAs wells. \magent{These phenomena} could arise from two mechanisms: (i) the Lorentz force from the emergent magnetic field due to injected electrons following the real-space spin texture and (ii) the induced emergent electric field (Faraday induction) arising from the time-dependent topological flux of the drifting PSL, which can drive the PSL lattice sideways~\cite{naga-toku}. Point (ii) is more likely to occur in our system{\blue{~\cite{non-Abelian}}}.
Recent experiments~\cite{yang, yang1} have successfully demonstrated the electrically-induced coherent propagation of helices in GaAs wells~\cite{G}. Further theoretical work similar to that in Ref.~\cite{Shen2014} is needed to fully describe the quantum transport properties of our PSL on top of a drifting Fermi sea, which can unveil topological phenomena in ordinary GaAs wells. 

\begin{acknowledgments}
We thank R. Raimondi, R. Winkler, K. Shen, A. Vishwanath, D. R. Candido and F. Zhang for useful discussions. This work was supported by FAPESP, CNPq, PRP/USP (Q-Nano), the Swiss NSF, NCCR QSIT, and the Natural Science Foundation of China (Grant No. 11004120). \red{ P.H.P. acknowledges CNPq support under the CsF program.}
\end{acknowledgments}


\begin{thebibliography}{30}
%
\bibitem{Sondhi} S. L. Sondhi, A. Karlhede, S. A. Kivelson, and E. H. Rezayi, Phys. Rev. B \textbf{47}, 16419 (1993).
%
\bibitem{Brey} L. Brey, H. A. Fertig, R. C\^ ot\' e, and A. H. MacDonald, Phys. Rev. Lett. \textbf{75}, 2562 (1995).
%
\bibitem{skyrmions} S. M\"uhlbauer \textit{et al.} Science \textbf{323}, 915 (2009); X. Z. Yu  \textit{et al.} Nature \textbf{465}, 901 (2010); S. Seki  \textit{et al.} Science \textbf{336}, 198 (2012).  See also M. B. A. Jalil and S. G. Tan, Sci. Rep. \textbf{4}, 5123 (2014). 
%
\bibitem{Overhauser} A. W. Overhauser, Phys. Rev. \textbf{128}, 1437 (1962).
%
\bibitem{naga-toku}For recent reviews, see N. Nagaosa and Y. Tokura, Nat. Nanotech. \textbf{8}, 899 (2013) and A. Fert, V. Cros, and J. Sampaio, Nat. Nanotech. \textbf{8}, 152 (2013).
%
\bibitem{Bernardes} E. Bernardes, J. Schliemann, M. Lee, J. C. Egues, and D. Loss, Phys. Rev. Lett. \textbf{99}, 076603 (2007). 
%
\bibitem{souma:2013} S. Souma, A. Sawada, H. Chen, Y. Sekine, M. Eto, and T. Koga, Phys. Rev. Applied {\bf 4}, 034010 (2015).
%
\bibitem{Schliemann} J. Schliemann, J. C. Egues, and D. Loss, Phys. Rev. Lett. {\bf 90}, 146801 (2003).
%
\bibitem{Bernevig-PSH} B. A. Bernevig, J. Orenstein, and S. C. Zhang, Phys. Rev. Lett. \textbf{97}, 236601, (2006).
%
\bibitem{ballistic-PSH} PSH patterns for 2DEGs in the ballistic regime were studied by M.H. Liu, K.W. Chen, S. H. Chen, and C.-R. Chang Phys. Rev. B {\bf 74}, 235322 (2006); see also Ming-Hao Liu, Ching-Ray Chang, and Son-Hsien Chen, {\it ibid.} {\bf 71}, 153305 (2005) for other ballistic spin textures. {In addition, a symmetry-based discussion of PSHs in hole gases can be found in Dollinger {\it et al.}, Phys. Rev. B {\bf 90} 115306 (2014). More recently, an interesting theoretical work by Kammermeier {\it et al.} [arXiv: 1606.08774] has investigated the possibility of PSHs in zincblende wells of general crystal orientation. } 
%
\bibitem{Koralek} J. D. Koralek, C. Weber, J. Orenstein, B. A. Bernevig, S. C. Zhang, S. Mack, and D. Awschalom, Nature \textbf{458}, 610 (2009).
%
\bibitem{Walser} M. P. Walser, C. Reichl, W. Wegscheider, and G. Salis, Nature Physics \textbf{8}, 757 (2012).
%
\bibitem{binz} B. Binz and A. Vishwanath, Physica B {\bf 403}, 1336 (2008).
%
\bibitem{Mishchenko} E. G. Mishchenko, A. V. Shytov, and B. I. Halperin, Phys. Rev. Lett. \textbf{93}, 226602 (2004). 
%
\blue{\bibitem{Raimondi2006} R. Raimondi, C. Gorini, P. Schwab, and M. Dzierzawa, Phys. Rev. B \textbf{74}, 035340 (2006).}
%
\bibitem{Duckheim} M. Duckheim, D.L. Maslov, and D. Loss, Phys. Rev. B \textbf{80}, 235327 (2009).
%
\blue{\bibitem{Wang} L. Y. Wang, C. S. Chu, and A. G. Mal'shukov,  Phys. Rev. B {\bf 81}, 115312 (2010).}
%
\bibitem{Sinova} X. Liu and J. Sinova, Phys. Rev. B \textbf{86}, 174301 (2012).
%
\bibitem{Shen2014} K. Shen, R. Raimondi, and G. Vignale, Phys. Rev. B {\bf 90}, 245302 (2014).
\red{\bibitem{suppl-mat-refs}See Supplemental Material for detailed descriptions about the SO interaction,
spin dynamics and spin relaxation, which includes Refs. \cite{Winkler,Rammer,Shen13,Meier,fu:2015,Raimondi2010,Sherman,pablo}.
 \bibitem{Winkler} R. Winkler, \textit{Spin-Orbit Coupling Effects in Two-Dimensional Electron and Hole Systems, Springer Tracts in Modern Physics}  Vol. 191 (Springer, New York, 2003). 
\bibitem{Rammer} J. Rammer, \textit{Quantum Field Theory of Nonequlibrium States }(Cambridge University Press, Cambridge, 2007).
\bibitem{Shen13} K. Shen and G. Vignale, Phys. Rev. Lett. \textbf{111}, 136602 (2013).
\bibitem{Meier} F. Meier and B.P. Zakharchenya, {\em Optical Orientation} (North-Holland, Amsterdam, 1984).
\bibitem{fu:2015} J. Y. Fu and J. C. Egues, Phys. Rev. B \textbf{91}, 075408 (2015).
\bibitem{Raimondi2010} R. Raimondi and P. Schwab, Physica E \textbf{42}, 952 (2010).
\bibitem{Sherman} E. Ya. Sherman, Appl. Phys. Lett. \textbf{82}, 209 (2003).
 \bibitem{pablo} P. I. Tamborenea, M. A. Kuroda, and F. L. Bottesi, Phys. Rev. B {\bf 68}, 245205 (2003).}
%
\bibitem{Calsaverini} R. S. Calsaverini, E. Bernardes, J. C. Egues, and D. Loss, Phys. Rev. B \textbf{78}, 155313 (2008).
%
\bibitem{real-param} The matrix elements $\eta$ and $\Gamma$ can be chosen real, see footnote [15] in Ref.~\cite{Bernardes}.
%
\bibitem{Dominik} F. Dettwiler, J. Y. Fu, P. J. Weigele, S. Mack, J. C. Egues, D. D. Awschalom, and D. Zumb\" uhl, arXiv:1403.3518.
%
\bibitem{note-pot} We assume $V(\mathbf{r})$ is the same for both subbands. 
%
\bibitem{Oestreich}S. D\"ohrmann, D. H\"agele, J. Rudolph,  M. Bichler,  D. Schuh,  and M. Oestreich,  Phys. Rev. Lett. {\bf 93}, 147405 (2004).
%
\bibitem{luffe:2011} M. C. L\"uffe, J. Kailasvuori, and T. S. Nunner, Phys. Rev. B \textbf{84}, 075326 (2011).
%
\bibitem{kiselev} This argument agrees with the numerical results of  Kiselev and Kim [Phys. Status Solidi (b) {\bf 221}, 491 (2000)].
%
%
%
\bibitem{kc-note} See Eq.~(S57) (SM) for $E_{\lambda_1,\lambda_2}\left(\mathbf{k}\right)$ with the particular choice $\Gamma=-\eta$; $k_c$ is defined in Eq.~(S28), SM. 
\bibitem{Bentmann} Some of these features resemble the data of Bentmann \textit{et al.} [Phys. Rev. Lett. \textbf{108}, 196801 (2012)].
%
%
%
%
%
%
%
%
%
%
\bibitem{yang} L. Yang,  \textit{et al.}, Nat. Phys. {\bf 8}, 153 (2012). 
%
\bibitem{SKHE} J. Zang, M. Mostovoy, J. H. Han and N. Nagaosa, Phys. Rev. Lett. {\bf 107}, 136804 (2011).
%
\blue{\bibitem{non-Abelian}Non-Abelian SO electromagnetic fields were discussed by Tokatly [Phys. Rev. Lett. {\bf{101}},106601 (2008)] and Gorini, {\it{et al.}} [Phys. Rev. B {\bf{82}}, 195316 (2010)]. For a more comprehensive review see Fujita, {\it{et al.}} [J. Appl. Phys.{ \bf{110}}, 121301 (2011)].}
%
\bibitem{yang1} L. Yang, \textit{et al.}, Phys. Rev. Lett. {\bf 109}, 246603 (2012).
%
\blue{\bibitem{G} See G. J. Ferreira \textit{et al.}, arXiv: 1608.05437, for a recent theoretical study of the spin drift and diffusion in spin-orbit coupled 2DEGs using a random walk model.}




\end{thebibliography}
\end{document}